\documentclass[12pt,preprint]{aastex}

\def\ms{m~s$^{-1}$}
\def\ks{km~s$^{-1}$}

\def\mjup{$M_{\rm Jup}$}
\def\rjup{$R_{\rm Jup}$}

\def\chis{$\chi^2_\nu$}

\def\feh{[Fe/H]}

\begin{document}

\title{Measurement of the Spin-Orbit Angle of Exoplanet HAT-P-1b\altaffilmark{1}}   

\author{
  John Asher Johnson\altaffilmark{2,3}, 
  Joshua N.\ Winn\altaffilmark{4}, 
  Norio Narita\altaffilmark{5},
  Keigo Enya\altaffilmark{6},
  Peter K.\ G.\ Williams\altaffilmark{2}, 
  Geoffrey W.\ Marcy\altaffilmark{2}, 
  Bun'ei Sato\altaffilmark{7},
  Yasuhiro Ohta\altaffilmark{8}, 
  Atsushi Taruya\altaffilmark{8},
  Yasushi Suto\altaffilmark{8},
  Edwin L.\ Turner\altaffilmark{9},
  Gaspar Bakos\altaffilmark{10},
  R. Paul Butler\altaffilmark{11},
  Steven S.\ Vogt\altaffilmark{12},
  Wako Aoki\altaffilmark{5}, 
  Motohide Tamura\altaffilmark{5},
  Toru Yamada\altaffilmark{13},
  Yuzuru Yoshii\altaffilmark{14},
  Marton Hidas\altaffilmark{15}
}

\email{johnjohn@ifa.hawaii.edu}

\altaffiltext{1}{Based on observations obtained at the Keck
  Observatory, which is operated as a scientific partnership among
  the California Institute of Technology, the University of
  California, and the National Aeronautics and Space Administration;
  the Subaru Telescope, which is operated by the National Astronomical
  Observatory of Japan;
  and the Lick Observatory, which is operated by the University of
  California.} 
\altaffiltext{2}{Department of Astronomy, University of California,
  Mail Code 3411, Berkeley, CA 94720}
\altaffiltext{3}{Current Address: Institute for Astronomy, University
  of Hawaii, Honolulu, HI 96822; NSF Postdoctoral Fellow} 
\altaffiltext{4}{Department of Physics, and Kavli Institute for
  Astrophysics and Space Research, Massachusetts Institute  
  of Technology, Cambridge, MA 02139}
\altaffiltext{5}{National Astronomical Observatory of Japan,
  2-21-1 Osawa, Mitaka, Tokyo 181--8588, Japan}
\altaffiltext{6}{Department of Infrared Astrophysics, Institute of
  Space and Astronautical Science, Japan Aerospace Exploration Agency,
  3-1-1 Yoshinodai, Sagamihara, Kanagawa 229--8510, Japan} 
\altaffiltext{7}{Global Edge Institute, Tokyo Institute of Technology,
2-12-1 Okayama, Meguro, Tokyo 152-8550, Japan}
\altaffiltext{8}{Department of Physics, School of Science, The
  University of Tokyo, 7-3-1 Hongo, Bunkyo-ku, Tokyo 113--0033, Japan}
\altaffiltext{9}{Princeton University Observatory, Peyton Hall,
Princeton, NJ 08544, USA}
\altaffiltext{10}{Harvard-Smithsonian Center for Astrophysics, 60
  Garden Street, Cambridge, MA 02138; NSF Postdoctoral Fellow}
\altaffiltext{11}{Department of Terrestrial Magnetism, Carnegie
  Institution of Washington DC, 5241 Broad Branch Rd. NW, Washington DC,
 20015-1305} 
\altaffiltext{12}{UCO/Lick Observatory, University of California at
  Santa Cruz, Santa Cruz, CA 95064} 
\altaffiltext{13}{Astronomical Institute, Tohoku University, Aramaki,
  Aoba, Sendai, 980-8578, Japan}
\altaffiltext{14}{Institute of Astronomy, School of Science, The
  University of Tokyo, 2-21-1 Osawa, Mitaka, Tokyo 181--0015, Japan} 
\altaffiltext{15}{Las Cumbres Observatory, 6740 Cortona Dr.~Suite 102,
  Santa Barbara, CA 93117}

\begin{abstract}
  We present new spectroscopic and photometric observations of the
  HAT-P-1 planetary system. Spectra obtained during three transits
  exhibit the Rossiter-McLaughlin effect, allowing us to measure the
  angle between the sky projections of the stellar spin axis and orbit
  normal, $\lambda = 3\fdg 7 \pm 2\fdg 1$. The small value of
  $\lambda$ for this and other systems suggests that the dominant
  planet migration mechanism preserves spin-orbit alignment. Using two
  new transit light curves, we refine the transit ephemeris and reduce
  the uncertainty in the orbital period by an order of magnitude. We
  find a upper limit on the orbital eccentricity of 0.067, with 99\%
  confidence, by combining our new radial-velocity measurements with
  those obtained previously.
\end{abstract}


\section{Introduction}

Prior to 1995, it was expected that Jovian planets around other stars
would inhabit wide, circular orbits similar to the Solar System gas
giants. It was therefore a surprise when the first exoplanet was
discovered with a minimum mass of 0.468~\mjup\ and a semimajor axis of
only 0.05~AU \citep{mayor95}. Since then, 85 ``hot Jupiters''---Jovian
planets with periods $\leq$10~days---have been detected around
Sun-like stars \citep{butler06,torres08}. It is unlikely that these
planets formed \emph{in situ}\, due to the low surface densities and
high temperatures of the inner regions of circumstellar disks
\citep{lin96}. A more likely scenario is that these massive planets
formed at a distance of several astronomical units, and then migrated
inward to their current locations.

Theories for the inward migration of planets can be divided into two
broad categories. The first category involves tidal interactions
between the planet and a remaining gaseous disk
\citep{lin96,moorhead08}. The second category involves few-body
gravitational dynamics, such as planet--planet scattering
\citep{rasio96,chat07}, dynamical relaxation \citep{pap01,adams03},
and Kozai cycles accompanied by tidal friction
\citep{holman97,fab07,wu07,nagasawa08}. One possible way to
distinguish between these categories is to examine the present-day
alignment between the stellar rotation axis and the planetary orbital axis.
Assuming that these axes were initially well aligned, disk-planet
tidal interactions would preserve this close alignment \citep{ward94},
while the second category of theories would at least occasionally
result in large misalignments. For example, \citet{adams03} predict a
final inclination distribution for dynamically relaxed planetary
systems that peaks near $20^\circ$ and extends to $85^\circ$.
Likewise, \citet{fab07} and \citet{wu07} simulated systems of planets
with randomly aligned outer companions and found that the Kozai
interaction resulted in a wide distribution of final orbital
inclinations for the inner planet, with retrograde orbits ($\lambda >
90^\circ$) not uncommon. Similar results were found by
\citet{nagasawa08}, for the case in which Kozai oscillations are
caused by an outer planet, rather than a companion star.

Spin-orbit alignment can be measured by taking advantage of the
Rossiter--McLaughlin (RM) effect that occurs during a planetary
transit. As the planet blocks portions of the rotating stellar
surface, the star's rotational broadening kernel becomes asymmetric
and its spectrum appears to be anomalously Doppler-shifted. The RM
effect has previously been observed and modeled for eight transiting
planetary systems \citep{queloz00,winn05,winn06b,winn07a,wolf07,
  narita07,narita08,bouchy08,lo08,winn08}. In this work, we add HAT-P-1 to
this sample.

HAT-P-1 (ADS\,16402B) is a member of a G0V/G0V visual binary and
harbors a short--period, Jovian planet. The transits of HAT-P-1b were
discoverd by \citet{bakos07} as part of the Hungarian-made Automated
Telescope Network (HATNet). The planet has a 4.465~day orbital period,
a mass of 0.53~\mjup, a radius $R_P = 1.20$~\rjup\ \citep{bakos07,
  winn07b}. We have monitored HAT-P-1 using precise radial velocity
(RV) and photometric measurements made both in and out of transit in
order to measure the RM effect and improve the precision with which
the system's orbital parameters are known. In the following section we
describe our observations and data reduction procedures. In~\S
\ref{model} we present the transit model that we fit to our
observations, and in \S~\ref{results} we present our results, and we
conclude in \S~\ref{discussion} with a brief discussion.

\section{Observations and Data Reduction}
\label{observations}
\subsection{Radial Velocity Measurements}
\label{sec:rv}

We observed the optical spectrum of HAT-P-1 using the High Resolution
Echelle Spectrometer \citep[HIRES,][]{vogt94} on the Keck~I 10m
telescope and the High Dispersion Spectrograph
\citep[HDS,][]{noguchi02} on the Subaru~8m telescope. We set up the
HIRES spectrometer in the same manner that has been used consistently
for the California-Carnegie planet search
\citep{butler96,marcy05b}. This is also the same setup that was used
to gather the 9 Keck/HIRES spectra reported by
\citet{bakos07}. Specifically, we employed the red cross-disperser and
used the I$_2$ absorption cell to calibrate the instrumental response
and the wavelength scale. The slit width was set by the $0\farcs85$ B5
decker, and the typical exposure times ranged from 3--5~min, giving a
resolution of about 60,000 at 5500\AA\ and a signal-to-noise ratio
(SNR) of approximately 120~pixel$^{-1}$. We gathered 3 spectra on
several nights when transits were not occurring, in order to refine
the parameters of the spectroscopic orbit. In addition we gathered a
dense time series of spectra on each of two nights, UT~2007~July~6 and
UT~2007~September~2, when transits were predicted to occur. On each
night we attempted to observe the star for many hours bracketing the
predicted transit midpoint, but there were interruptions due to clouds
and pointing failures. However, both nights of data provide good phase
coverage of the entire transit event. In total we obtained 79 new
Keck/HIRES spectra, of which 49 were observed while a transit was
happening.

For our Subaru/HDS spectra we employed the standard I2a setup of the
HDS, covering the wavelength range 4940--6180\AA\ with the I$_2$
absorption cell. The slit width of $0\farcs 8$ yielded a spectral
resolution of $\sim$45,000. The typical exposure time was 10~min
resulting in a SNR of 120~pixel$^{-1}$. Our Subaru observations took
place on 3 different nights spread out over 2 months. Two of the
nights were not transit nights; we gathered 8 spectra on those nights
in order to refine the parameters of the spectroscopic orbit. The last
night, UT~2007~September~20, was a transit night, and we gathered 25
spectra over 7.3~hr bracketing the predicted transit midpoint, of
which 16 were gathered during the transit.

\begin{figure}[p]
\epsscale{1.0}
\plotone{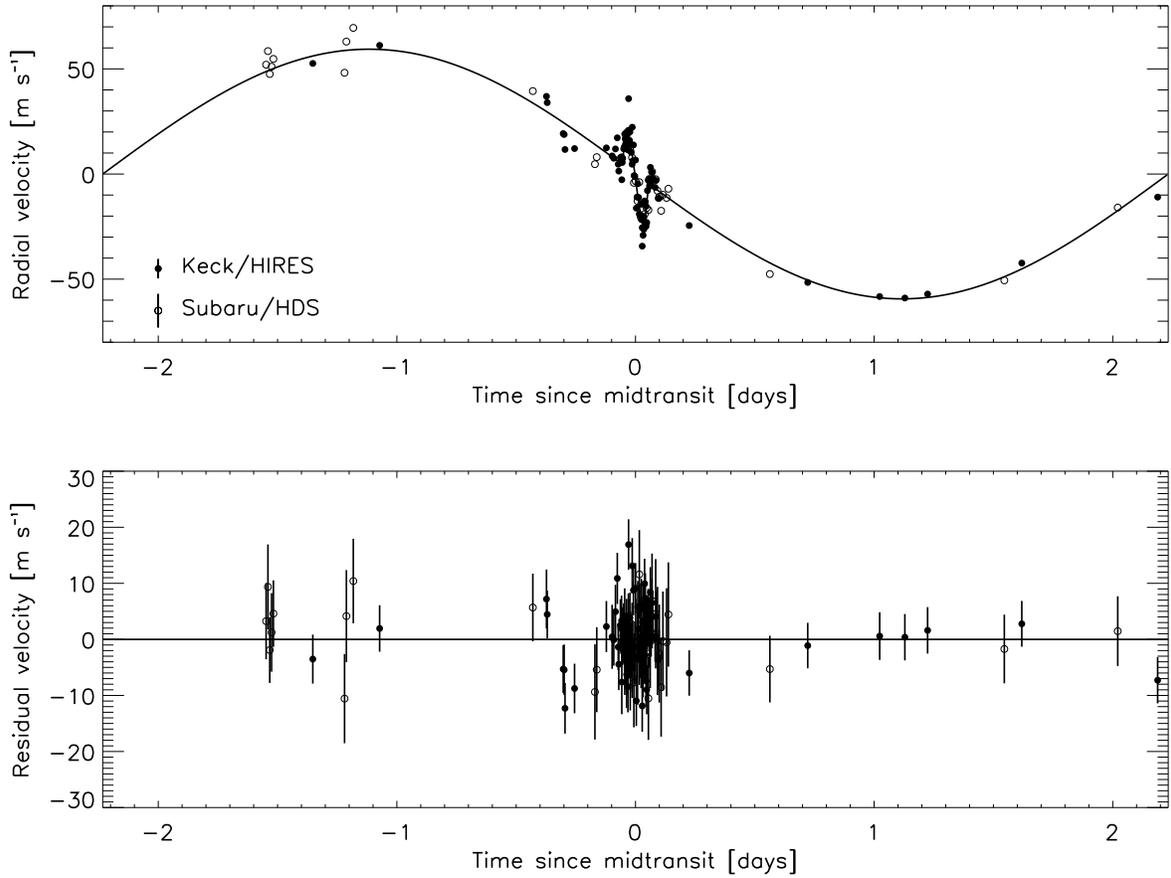}
\caption{{\it Top:} Relative radial velocity measurements of HAT-P-1,
  from this work and from 
Bakos et al.~(2007). The solid line is the best-fitting model. The
typical measurement uncertainties are illustrated as points with error
bars in the lower left corner. A detailed view near midtransit is
shown in Figure~\ref{fig:transit}. {\it Bottom:} Residual radial
  velocities after subtracting the best-fitting model.
\label{fig:rv}}
\end{figure}

\begin{figure}[p]
\epsscale{0.8}
\plotone{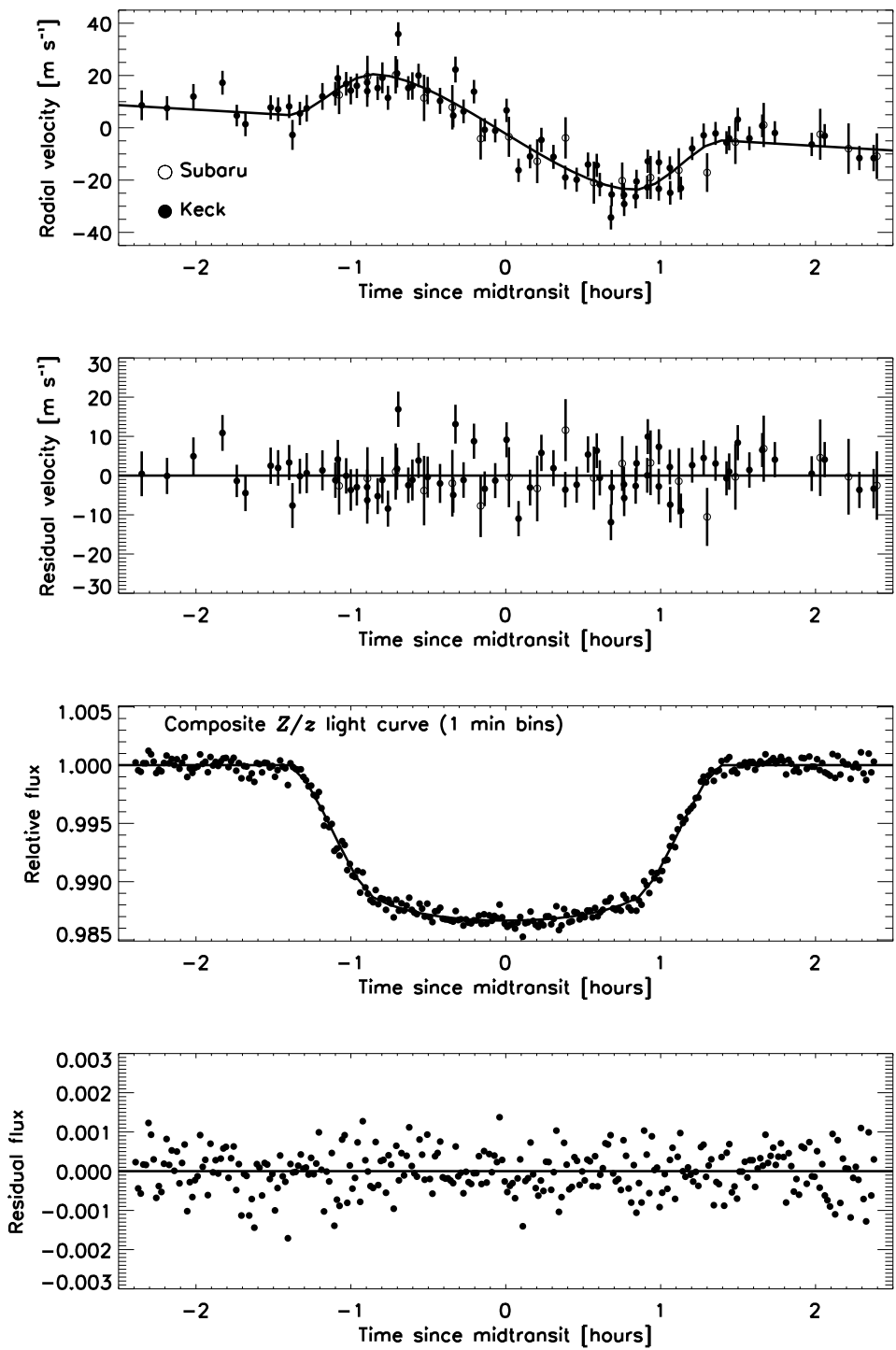}
\caption{ 
{\it Top:} Relative radial velocity measurements of HAT-P-1,
centered on the midtransit time.
{\it Bottom:} Composite, time-binned transit light curve of HAT-P-1,
based on the $Z$ and $z$-band data from Winn et al.~(2007) and this
work. The rms scatter of the residuals is 0.00057. This composite
light curve was fitted simultaneously with the radial-velocity data.
\label{fig:transit}}
\end{figure}

We performed the Doppler analysis with the algorithm of
\citet{butler96}. For the Subaru data we used a version of this
algorithm customized for HDS by \citet{sato02}. We estimated the
measurement error in the Doppler shift derived from a given spectrum
based on the weighted standard deviation of the mean among the
solutions for individual 2~\AA\ spectral segments. The typical
measurement error was 3~m~s$^{-1}$ for the Keck data and 7~m~s$^{-1}$
for the Subaru data. The data are given in Table~\ref{tab:vels} and
plotted in Figs.~\ref{fig:rv} and \ref{fig:transit}. Also given in
that table, and shown in those figures, are data based on the 9
Keck/HIRES spectra and 4 Subaru/HDS spectra obtained previously by
\citet{bakos07}. We note that the RV timestamps reported by Bakos et
al.~(2007) are incorrect. They were said to be Heliocentric Julian
dates, but they are actually Julian dates. We provide the corrected
dates in Table~\ref{tab:vels}.

\subsection{Photometric Measurements}

We obtained photometric measurements of HAT-P-1 during the transit of
UT~2007~Oct~8 using the Nickel 1m telescope at Lick Observatory on
Mount Hamilton, California. We used the Nickel Direct Imaging Camera,
which is a thinned Loral $2048^2$ CCD with a $6.3\arcmin$ square field
of view\footnote{This is the same camera used by \citet{winn07b},
  which they mistakenly described as a 2048$^2$ Lawrence Labs CCD with
  a $6\farcm1 \times 6\farcm1$ field of view.}. We observed through a
Gunn $Z$ filter, and used $2\times2$ binning for an effective pixel
scale of $0\farcs37$~pixel$^{-1}$. The exposure times varied depending
upon conditions but were typically 10-12~s, with a readout and setup
time between exposures of 34~s. The conditions were clear for most of
the transit with $\sim1\farcs0$ seeing. However, observations during
ingress were partially obscured by clouds and the data from that time
period proved to be significantly noisier than the rest; we have
excluded those data from our analysis. We determined the instrumental
magnitude of HAT-P-1 relative to two comparison stars using an
aperture with an 11 pixel radius and a sky background annulus
extending from 15 to 18 pixels.

We observed the transit of UT~2007~September~20 with the MAGNUM~2m
telescope on Haleakala, in Hawaii
\citep{kobayashi98,yoshii02,yoshii03}. The MAGNUM photometric
observations were conducted on the same night as the Subaru/HDS
transit observations described in \S~\ref{sec:rv}. We employed the
Multicolor Imaging Photometer (MIP), using a $1024^2$ SITe CCD with a
pixel scale of $0\farcs277$~pixel$^{-1}$. The camera's field of view
is $1\farcm5$, which is much smaller than the field of view of the
detector. During each exposure, the field was shifted on the detector
along a $3\times3$ grid, which allowed us to increase the duty cycle
since the chip was read out only once for every 9
exposures. Observations were made through a Johnson $V$-band filter,
and the exposure times were 10~s, with 40~s per exposure for readout
and setup. The MIP images were reduced with the standard pipeline
described by \citet{minezaki04}. We determined the instrumental
magnitude of HAT-P-1 relative to its visual binary companion,
ADS\,16402A, using an aperture radius of 15 pixels, and estimated the
sky background level with an annulus from 20 to 25 pixels.

\begin{figure}[p]
\epsscale{1.0}
\plotone{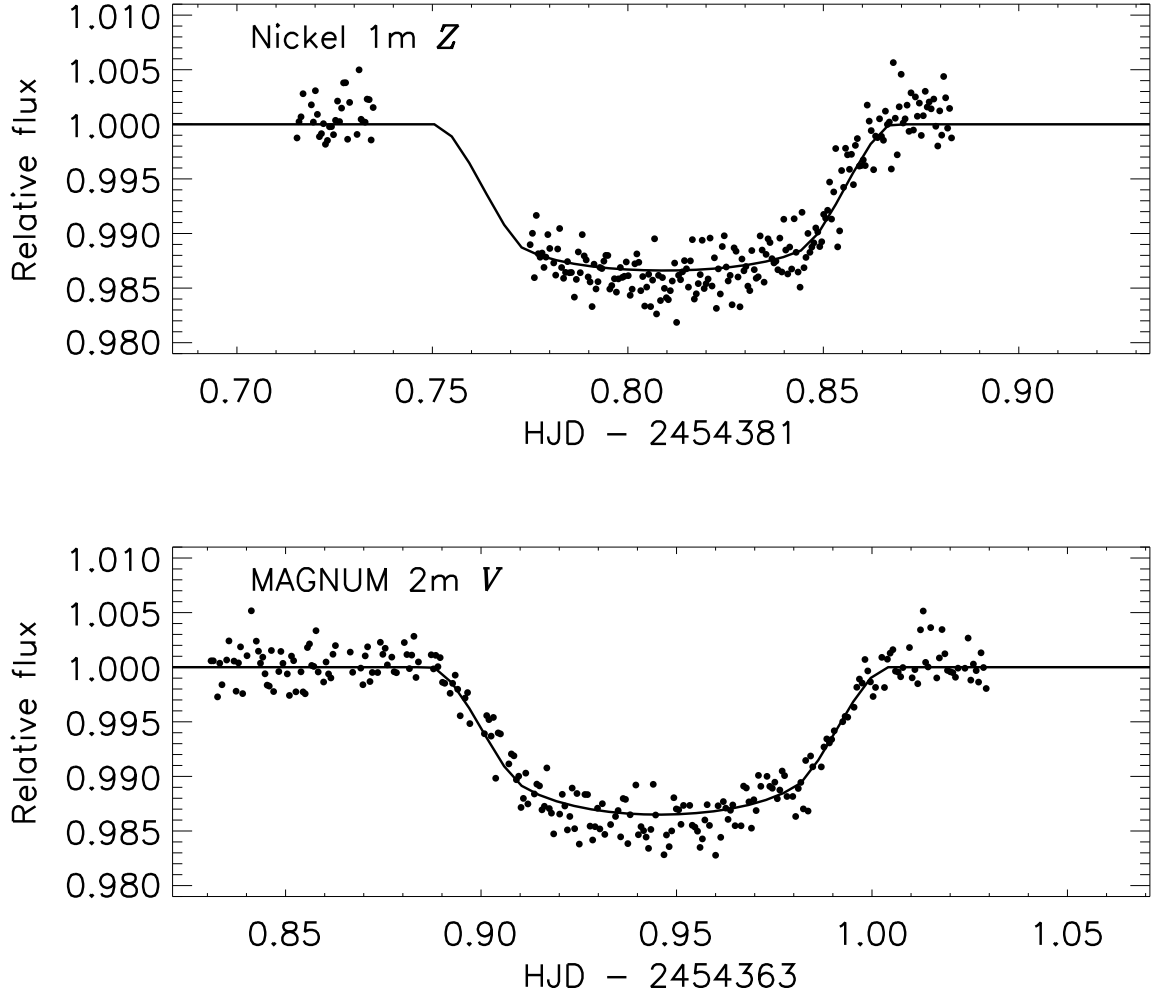}
\caption{ 
Photometry of transits of HAT-P-1, using the Nickel~1m telescope
and a $Z$-band filter (top), and the MAGNUM~2m telescope and a
$V$-band filter (bottom). The solid lines show the best-fitting model.
\label{fig:phot}}
\end{figure}

The photometric data are given in Table~\ref{tab:phot} and plotted in
Fig.~\ref{fig:phot}.  In the final light curves, the root-mean-squared
(rms) relative flux, outside of transits, is 0.0019 for the Nickel
data and 0.0016 for the MAGNUM data.

\section{The Model}
\label{model}
\subsection{An Updated Ephemeris}
\label{newphot}

The extended time baseline of our new photometric measurements allows
us to refine the transit ephemeris. We first computed midtransit times
from the light curves using the method described by
\citet{winn07b}. In particular, to assign proper weights to the
photometric data during the light curve fitting procedure, we applied
a correction to the uncertainties to take into account time-correlated
noise (``red noise''), which was determined by examining the rms
residuals in time-averaged light curves (see \citet{winn07b}). This
resulted in a factor-of-two increase in the error bars, relative to a
situation in which correlated noise is ignored.

We employed the same modeling procedure described in detail by
\citet{holman06} and \citet{winn07} and summarized as follows. We
modeled the path of the planet across the stellar disk using a
parameterized model based on a planet and star in a Keplerian orbit
about their center of mass. We fitted the photometric observations
using the analytic formulas of \citet{mandelagol} and a quadratic
limb--darkening law with fixed coefficients a based on the tabulated
calculations of \citet{claret04}.\footnote{For the $Z$ band, the
  coefficients were $a_Z=0.18$ and $b_Z=0.34$. For the $V$ band, the
  coefficients were $a_V=0.40$ and $b_V=0.32$.} The free parameters
were the scaled stellar radius $R_\star/a$ (where $a$ is the semimajor
axis), the planet-to-star radius ratio $R_p/R_\star$, the orbital
inclination $i$; and for each light curve, the midtransit time $T_c$,
the mean out-of-transit flux, and a time gradient of the
out-of-transit flux (to account for some systematic errors in the
photometry). The model fit was carried out using a Markov Chain Monte
Carlo (MCMC) algorithm with $10^6$ links, in which a single
randomly-chosen parameter was perturbed at each link, with a
perturbation size tuned such that $\sim$40\% of the jumps were
executed. The mean values and standard deviations of the posterior
probability distributions (which were nearly Gaussian in this case)
were adopted as the ``best--fit'' parameters and uncertainties.

\begin{figure}[p]
\epsscale{1.0}
\plotone{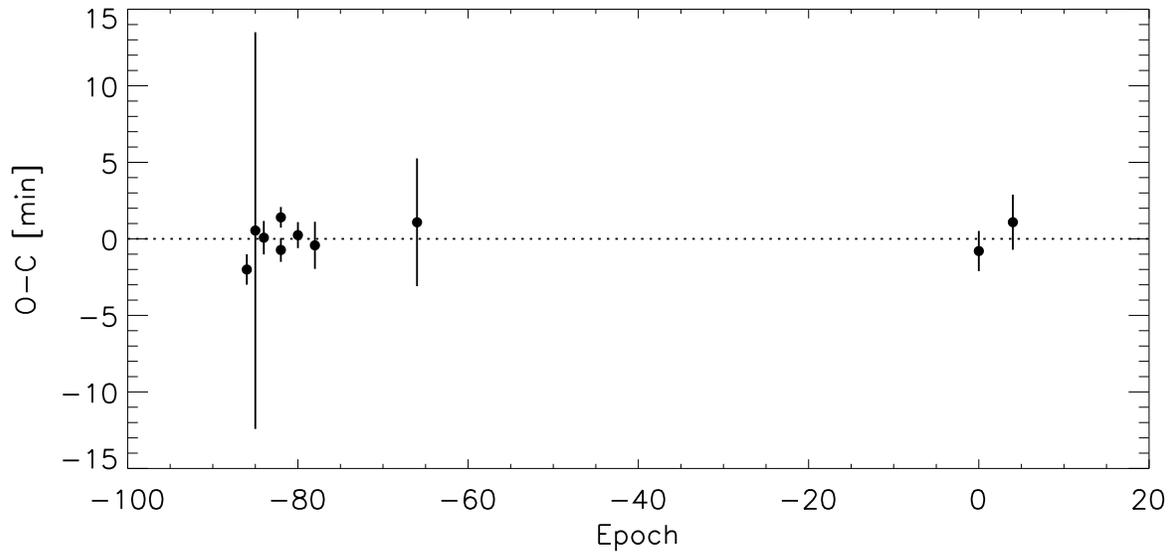}
\caption{ 
Residuals from a linear ephemeris that was fitted to the transit times in
Table~\ref{tab:tc}. The best--fitting ephemeris has a period $P =
4.4652934 \pm 0.0000093$~days and midtransit time 
$T_c = 2454363.94656 \pm 0.00072$ (HJD).
\label{fig:tc}}
\end{figure}

We fit a linear ephemeris to all of the times listed in
Table~\ref{tab:tc}, which includes the new transit times and those
measured by \citet{winn07b}. We found that two of the entries in
Table~3 of \citet{winn07b} were incorrect: the first time was wrong
because the data had not been normalized correctly, and the sixth time
was too small by one period because of a rounding error in the
computer code that generated the table. The corrected times are given
in Table~\ref{tab:tc}. A linear fit to the transit times had
$\chi^2=10.2$ and 8 degrees of freedom, indicating an acceptable
fit. The fit residuals are plotted in Figure~\ref{fig:tc} and the
updated ephemeris is given in Table~\ref{tab:params}. The uncertainty
in our updated period is about 10 times smaller than the previous
estimate. In our subsequent analysis we fix the period at this value,
as the uncertainty is negligible for our purposes.

\subsection{The Orbital Eccentricity}

\citep{bakos07} reported a tentative detection of a nonzero orbital
eccentricity, $e=0.09 \pm 0.02$, based on an analysis of 13 RV
measurements. With our expanded RV data set, we can check on this
tentative detection. We modeled our radial velocity measurements using
a Keplerian orbit with 6 free parameters: the velocity semiamplitude
$K$, the orbital eccentricity $e$, the argument of pericenter
$\omega$, and an additive velocity for each of the 3 velocity groups
(our Keck velocties and those of \citet{bakos07}; our Subaru
velocities; and the Subaru velocities of \citet{bakos07}). The time of
transit, $T_c$, and the orbital period $P$ were held fixed at the
values determined from the photometric data.

To avoid complications at this stage due to the RM effect, we fitted
only those 43 velocities that were gathered well outside of
transits. Specifically we excluded all velocities that were measured
within a 4~hr window centered on the calculated midtransit time (the
actual transit duration is 2.8~hr). To assign proper weights to the RV
measurements we needed to estimate the noise due to astrophysical
sources such as stellar pulsation or rotational modulation of surface
features, commonly known as ``jitter'' \citep{saar98, wright05}. We
found it necessary to add (in quadrature) 3.7~\ms\ to the measurement
errors in order to obtain a \chis\ of unity. This jitter estimate is
consistent with the 3.4~\ms\ predicted by \citet{wright05} and used by
\citet{bakos07}. In the modeling procedure described in the rest of
this section, we used the augmented error bars, while
Table~\ref{tab:vels} gives only the internal measurement
uncertainties. Unlike our previous analyses of HD~189733 and
HD\,147506 \citep{winn06b,winn07a}, we found no evidence for a higher
night--to--night jitter compared to the intra--night jitter. We
therefore did not modify the error bars any further than the
quadrature addition of our jitter estimate.

We employed an MCMC fitting algorithm using $10^6$ steps and
perturbation sizes resulting in a 30-50\% acceptance rate
\citep[e.g.][]{winn05}. The orthogonal parameters describing the
eccentricity $e$ and argument of periastron $\omega$ were $e\cos\omega
= 0.003\pm 0.013$, $e\sin\omega = 0.004\pm 0.025$. The orbital
eccentricity of the HAT-P-1 system was found to be smaller than 0.067
with 99\% confidence. This is consistent with the theoretical
expectation that the orbit should have circularized due to tidal
friction. The circularization timescale is $\sim$0.23~Gyr assuming a
tidal quality factor of $10^6$ \citep{bakos07}, and the estimated
stellar age is 2.7~Gyr \citep{torres08}. In what follows we assume
$e=0$ exactly.

\subsection{Joint Analysis of Radial Velocities and Photometry}
\label{subsec:joint}

To determine the projected spin-orbit angle and its uncertainty, we
simultaneously fitted a parametric model to the RV data as well as a
composite transit light curve, generated from all of the $Z$ and $z$
photometric data at our disposal, from this work and from
\citet{winn07b}. The composite light curve has 1 minute bins, and an
out-of-transit rms of 0.00057. It is shown in Fig.~\ref{fig:transit}
along with the transit RVs. Although our main interest is in the
spin-orbit parameters, which are largely determined by the transit RV
data, we included the photometric data in the fit as a convenient way
to account for the uncertainties in the photometric parameters and
their covariances with the spin-orbit parameters, although in practice
these covariances proved to be small.

The aspects of the model that attempt to fit the photometry, and the
orbital Doppler data, have already been described. To calculate the
radial velocity during transits, we must calibrate the relationship
between the ``anomalous Doppler shift'' that is returned by our code
for measuring Doppler shifts, and the physical parameters and
configuration of the star and planet. For this purpose we used the
technique of \citet{winn05}, in which simulated stellar spectra are
created that exhibit the RM effect, and then these spectra are
analyzed with the same Doppler-measuring code that is used on actual
data. Such simulations are needed because the algorithm for measuring
Doppler shifts involves fitting for parameters that are intended to
describe the time-variable instrumental profile of the spectrograph,
and these parameters may interact with the spectral disortion of the
RM effect in ways that are hard to predict.

In our simulations the physical configuration of the planet and star
is characterized by the transit flux decrement, $\epsilon$, and the
velocity of the occulted portion of the stellar disk (the
``sub--planet velocity''), denoted by $v_p$. We created simulated
spectra with the same data format and noise characteristics as the
observations, and analyzed these with the same analysis pipeline used
for the actual observations. The simulated in--transit spectra are
based on a ``template'' spectrum representing the disk--integrated
spectrum of the star (described below), which we broaden to match the
rotational velocity of HAT-P-1.\footnote{The broadening kernel depends
  on the assumed limb-darkening law of the star, which we took to be a
  linear law with limb-darkening parameter $u=0.67$. The results for
  the RM calibration formula are insensitive to the choice of $u$;
  very similar results were obtained for the choices $u=0.2$ and
  $u=0.8$.} We subtract from this template spectrum an unbroadened
copy that is scaled by $\epsilon$ and Doppler--shifted by $v_p$, and
then measure the radial velocity anomaly $\Delta v$. We repeat this
process for a grid of $\{\epsilon, v_p\}$ and approximate $\Delta
v(\epsilon, v_p)$ with a two--dimensional polynomial fit. Differential
rotation was ignored, as its effects are expected to be negligible
\citep{gaudi07}.

The template spectrum should be similar to that of HAT-P-1 but with
narrower lines because of the lack of rotational broadening exhibited
by the sub--planet spectrum. We selected the NSO solar atlas
\citep{kurucz84} and a Keck/HIRES spectrum of HD\,34411
\citep[$T_{eff} = 5911$~K, \feh~$=+0.12$;][]{valenti05}.  We
found that the results based on either template are consistent with
the function $\Delta v = - \epsilon v_p$. This function is consistent
with the analytic expressions of \citet{ohta05} and
\citet{gimenez06}. We have found the best functional form of this ``RM
calibration'' to vary from system to system; we also found a linear
relation for TrES-2 (Winn et al.~2008). Thus, for this study, the
calculated radial velocity of the star was taken to be the sum of the
radial velocity of the Keplerian orbit, and the anomalous velocity
$\Delta v = -\epsilon v_p$.

The model had 12 free parameters: $K$, $R_\star/a$, $R_p/R_\star$,
$i$, $T_c$, $v\sin i_\star$, $\lambda$, three additive constants for
the 3 different groups of velocity data, and two limb-darkening
coefficients $a_Z$ and $u$ (to be explained in the next few
paragraphs, see also Table~\ref{tab:params}). The orbital period was
fixed at the value determined 
previously and the eccentricity was fixed at $e=0$. The two model
parameters relating to the RM effect are the 
line-of-sight stellar rotation velocity ($v \sin i_\star$), and the
angle between the projected stellar spin axis and orbit normal
($\lambda$). The projected spin-orbit angle $\lambda$ is measured
counterclockwise on the sky from the projected stellar rotational
angular-momentum vector to the projected orbital angular-momentum
vector (see Ohta et al.~2005 or Gaudi \& Winn 2007 for a diagram). Due
to the symmetry of the situation, a configuration with inclination $i$
and spin-orbit angle $\lambda$ cannot be distinguished from a
different configuration with inclination $180\arcdeg - i$ and
spin-orbit angle $-\lambda$. To break this degeneracy we restrict $i$
to the range from zero to 90 degrees, and allow $\lambda$ to range
from $-180\arcdeg$ to $+180\arcdeg$.

For the photometric model, the limb-darkening law was assumed to be
quadratic, $I/I_0 = 1 - a_Z(1-\mu) - b_Z(1-\mu)^2$, where $\mu$ is the
cosine of the angle between the line of sight and the local surface
normal. Given the precision of our data it is not possible to place
meaningful constraints on both $a_Z$ and $b_Z$. We fixed $b_Z=0.34$,
based on interpolation of the tables by \citet{claret04}. We allowed
$a_Z$ to be a free parameter subject to a mild {\it a priori}\,
constraint, shown in Eq.~(\ref{eq:chisqr}) below, that enforces
agreement with the tabulated value within $\approx$0.2. The choice of
0.2 is somewhat arbitrary and is fairly conservative, in the sense
that better agreement is usually observed between fitted and
theoretical limb darkening coefficients for the cases when such
comparisons can be made (see, e.g., Winn, Holman, \& Roussanova 2007
and Southworth~2008). This approach is intermediate between the
extreme approaches of fixing the limb darkening parameters exactly
(placing too much trust in tabulated values) and allowing them to be
completely free parameters (disregarding all theoretical knowledge of
stellar atmospheres and possibly allowing unphysical parameter
values).

For the RM model, we adopted a linear law [$I/I_0 = 1-u(1-\mu)$] for
simplicity, since a quadratic law does not seem justified by the
precision of the RM data. The appropriate choice of the limb-darkening
coefficient $u$ is not obvious. The Doppler-shift measurement is based
on the portion of the spectrum between 5000 and 6200~\AA, where the
iodine absorption lines are plentiful. The tables of \citet{claret04}
lead to an expectation $u\approx 0.67$ for this spectral
region. However, the Doppler information arises primarily from the
steep sides of the stellar absorption lines in this region, and the
degree of limb darkening in the lines may differ from the degree of
limb darkening in the continuum, since the line radiation arises from
a different depth in the stellar atmosphere.

To investigate this issue we examined a Kurucz (1979) ATLAS12
plane-parallel model stellar atmosphere with $T_{\rm eff} = 5750$~K,
$\log g=4.5$, \feh~$=0.0$)\footnote{Downloaded from {\tt
    kurucz.harvard.edu}.}, which was originally computed for the star
XO-1 (McCullough et al.~2007) but whose properties are a reasonable
match to those of HAT-P-1 ($T_{\rm eff} = 5975$~K, $\log
g=4.5$, \feh~$=+0.1$; Torres et al.~2008). The stellar intensity was
computed for 17 different values of $\mu$, with a resolving power of
500,000. When the spectrum is averaged between 500--620~nm, the
best-fitting linear limb-darkening coefficient is 0.65, in agreement
with \citet{claret04}. According to the model, the degree of limb
darkening is smaller in the steepest portions of absorption lines. For
example, for a band centered on one of the Mg~b triplet lines at
518.5~nm, we find $u=0.66$ when the bandwidth is $\Delta\lambda=1$~nm,
and $u=0.50$ when $\Delta\lambda=0.02$~nm (encompassing only the
steepest portion of the line). For other strong lines we also
find that $u$ is decreased by 0.1--0.2 in the cores. For this reason,
we chose to allow $u$ to be a free parameter, with the same type of
{\it a priori}\, constraint used for the photometric limb-darkening
law (see below).

The fitting statistic was
\begin{eqnarray}
\label{eq:chisqr}
\chi^2 & = &
\sum_{j=1}^{287}
\left[
\frac{f_j({\mathrm{obs}}) - f_j({\mathrm{calc}})}{\sigma_{f,j}}
\right]^2
+
\sum_{j=1}^{125}
\left[
\frac{v_j({\mathrm{obs}}) - v_j({\mathrm{calc}})}{\sigma_{v,j}}
\right]^2
+
\left(
\frac{a_Z-0.18}{0.2}
\right)^2
+
\left(
\frac{u-0.67}{0.2}
\right)^2
,
\end{eqnarray}
where $f_j$(obs) are the relative flux data from the composite light
curve and $\sigma_{f,j}$ is the out-of-transit rms.  Likewise
$v_j$(obs) and $\sigma_{v,j}$ are the radial-velocity measurements and
uncertainties after adding the jitter as described above. The last two
terms represent {\it a priori}\, constraints on the linear
limb-darkening coefficients $a_Z$ (for the photometric data) and $u$
(for the radial-velocity data). As before, we solved for the model
parameters and uncertainties using a Markov Chain Monte Carlo
algorithm. We used a chain length of $10^6$ steps and adjusted the
perturbation size to yield an acceptance rate of $\sim$40\%. The
posterior probability distributions for each parameter were roughly
Gaussian, so we adopt the mean as the ``best--fit'' value and the
standard deviation as the 1-$\sigma$ error. For the joint model fit
the minimum $\chi^2$ is 412.2, with 402 degrees of freedom, giving
$\chi^2_\nu = 1.03$.  This nearly ``perfect'' goodness-of-fit
statistic should be interpreted as a check on the appropriateness of
our data weights, rather than an independent check on the validity of
the model, because we inflated the RV errors and attributed the RV
scatter to jitter, and likewise we set the flux uncertainties equal to
the out-of-transit RMS flux.

\section{Results}
\label{results}

The results from our analysis are given in Table~\ref{tab:params}.
The parameters depending on the RV data are $v\sin i_\star = 3.74 \pm
0.30$~\ks\ and $\lambda = 3\fdg 7 \pm 2\fdg 1$. (Below, we
argue that the true error in $v\sin i_\star$ is subject to an
additional systematic error of 0.5~\ks.) The small $\lambda$
indicates close alignment between the sky--projected stellar spin axis
and orbit normal. Our measured rotation velocity is higher than the
value $v\sin i_\star = 2.2 \pm 0.2$~\ks that was reported by
\citet{bakos07}. Using the Spectroscopy Made Easy (SME) program
(Valenti \& Piskunov 1996, Valenti \& Fischer 2005), we reanalyzed the
same HIRES template observation of HAT-P-1 that was analyzed by Bakos
et al., and found $v\sin i_\star = 3.4 \pm 0.5$~\ks, which agrees with
the result of our RM analysis. All other spectroscopic parameters from
our SME analysis agreed with those reported previously. The primary
difference in our analysis is that we used an appropriately narrow
instrumental profile width, or equivalently a higher resolution, as
measured from the model instrumental profile used in our Doppler
analysis. The lower resolution assumed by Bakos et al.\ artificially
compensated for rotational broadening, resulting in an erroneously low
value of $v\sin i_\star$ (Debra~Fischer private communication 2008).

The parameters that rely primarily on our photometry agree well with
those of \citet{winn07b}, \citet{southworth07} and \citet{torres08};
and those authors generally agree with one another, though there are
differences in the exact treatment of red noise and limb darkening.
As an additional check on our quoted errors, and in particular our
assumption that the composite light curve had uncorrelated photometric
errors, we applied the ``residual-permutation'' or ``rosary-bead''
method. In this method, one calculates the residuals between the
photometric data and the best-fitting model, and creates many
different ``realizations'' of the data that preserve any
time-correlated noise by time-shifting the residuals and adding them
back to the model. Then, the {\it a posteriori}\, probability
distribution for each parameter is estimated by minimizing $\chi^2$
for each different realization of the data, and creating histograms of
the parameter values. The error bars returned by this method were
similar to, or smaller than, the error bars quoted in Table~4.

The result for the photometric limb-darkening parameter is $a_Z=0.26
\pm 0.06$, showing that the data prefer a slightly more limb-darkened
star than in the ATLAS models from which limb-darkening coefficients
were tabulated by \citet{claret04}. The result for the radial-velocity
limb-darkening parameter is $u=0.90_{-0.20}^{+0.03}$. This is about
1$\sigma$ {\it larger}\, than the expected continuum value of 0.67,
even though the ATLAS models predict that the steepest portion of the
absorption lines (which provide most of the Doppler information)
should exhibit {\it smaller}\, limb darkening. Assuming the models are
correct, it is possible that the limb-darkening parameter $u$ is
compensating for an inaccuracy in our model of the RM effect,
especially for the ingress and egress phases where limb darkening is
strongest. This issue deserves further investigation, perhaps by
increasing the sophistication of our RM calibration procedure (see
\S~\ref{subsec:joint}), using spatially resolved theoretical intensity
distributions rather than an empirical stellar template. Fortunately
this issue affects only the results for $v\sin i_\star$, and not for
$\lambda$. This can be understood because $u$ and $v\sin i_\star$ both
depend on the amplitude of the anomalous Doppler shift, while
$\lambda$ depends almost entirely on the timing of the null of the
anomalous Doppler shift.\footnote{For other transiting systems with
  small impact parameters, such as TrES-1 (Narita et al.~2007) and
  HAT-P-2 (Winn et al.~2007, Loeillet et al.~2008), there is a strong
  degeneracy between $v\sin i_\star$ and $\lambda$ (Gaudi \& Winn
  2007). This type of systematic error may affect the results for
  $\lambda$ in those cases.} We verified this by fixing $u$ at values
between 0.5 and 0.9 and observing that the results for $v\sin i_\star$
change by 0.5~km~s$^{-1}$ while the results for $\lambda$ are
unchanged. Thus, we conclude that our result for $v\sin i_\star$ is
subject to a systematic error of approximately 0.5~km~s$^{-1}$. In
Table 4, we have added this systematic error in quadrature to the
statistical error of 0.30~km~s$^{-1}$ giving a total error of
0.58~km~s$^{-1}$.

\section{Summary and Discussion}
\label{discussion}

We have obtained high--precision photometric and spectroscopic
measurements of the star HAT-P-1. Our in--transit spectroscopic
observations clearly show the anomalous Doppler shift due to the
Rossiter--McLaughlin effect, and we find that the angle between the
sky projections of the stellar spin axis and the orbit normal is
$3\fdg7\pm 2\fdg1$. Additional Doppler measurements made during
non-transit orbital phases allow us to constrain the orbital
eccentricity to $e < 0.067$ with 99\% confidence. We measured the
transit times from two new light curves and refined the orbital period
by nearly an order of magnitude.

The HAT-P-1 system is an interesting case for planetary migration
theories because it is known to have a stellar companion in a wide
orbit \citep{bakos07}, a key ingredient for the Kozai mechanism. It is
also suggestive that the radius of the planet HAT-P-1b is on the high
end of theoretical expectations \citep{bakos07,winn07b}, which may 
a relic of tidal energy dissipation \citep{fab07}. Had the planetary
system exhibited a large spin--orbit misalignment, it would have
provided evidence for a scenario in which HAT-P-1b migrated to its
current orbit as a result of Kozai oscillations, coupled with tidal
dissipation within the planetary interior \citep{fab07}. However, the
small value of $\lambda$ does not necessarily rule out the Kozai
migration scenario, as small spin--orbit angles are not excluded by
the simulations \citep{fab07, wu07, nagasawa08}. But taken together
with the 7 other planetary systems with small values of $\lambda$, it
seems likely that the dominant migration mechanism responsible for hot
Jupiters preserves spin--orbit alignment.

Even if gravitational few-body mechanisms such as Kozai cycles do not
represent the dominant migration channel for the formation of hot
Jupiters, these mechanisms may be nonetheless be responsible for
configurations of some of close--in planets. A prime example is the
orbit of HD\,17156b, for which \citet{narita08} reported a 2.5$\sigma$
detection of a large misalignment ($\lambda = 62^\circ \pm
25^\circ$). Since small values of $\lambda$ are not excluded by any of
the existing theories of planet migration, misaligned systems like
HD\,17156 provide the most important tests of the various migration
mechanisms. With wide--field transit surveys discovering new
planets at an accelerating pace, the sample of measured spin--orbit
angles will soon be large enough and precise enough to directly
confront the theory.

\acknowledgements We thank the students of GWM's Ay120 Advanced
Astronomy Lab course for observing and measuring the transit light
curve of HAT-P-1. In particular, we acknowledge the efforts of
Kimberly Aller, Niklaus Kemming, Anthony Shu and Edward Young. We
thank the UCO/Lick technical staff for the new remote-observing
capability, allowing the photometry to be carried out from UC
Berkeley. We are grateful for support from the NASA Keck PI Data
Analysis Fund (JPL 1326712). JAJ and GB are NSF Astronomy and
Astrophysics Postdoctoral Fellows with support from the NSF grant
AST-0702821. We appreciate funding from NASA grant NNG05GK92G (to
GWM). PKGW is supported by an NSF Graduate Research Fellowship. This
research has made use of the SIMBAD database operated at CDS,
Strasbourg, France, and the NASA ADS database. The authors wish to
extend special thanks to those of Hawaiian ancestry on whose sacred
mountain of Mauna Kea we are privileged to be guests. Without their
generous hospitality, the Keck and Subaru observations presented
herein would not have been possible.


\clearpage

\begin{deluxetable}{llll}
\tablecaption{Doppler Shift Measurements of HAT-P-1\label{tab:vels}}
\tabletypesize{\scriptsize}
\tablewidth{0pt}
\tablehead{
\colhead{Telescope Code\tablenotemark{a}} &
\colhead{Heliocentric Julian Date} &
\colhead{Radial Velocity} &
\colhead{Measurement Uncertainty} \\
\colhead{} &
\colhead{} &
\colhead{(m~s$^{-1}$)} &
\colhead{(m~s$^{-1}$)} 
}
\startdata
  1 &  $  2453927.06954$  &  $ -50.33$  &  $   1.66$  \\
  1 &  $  2453927.96670$  &  $ -41.12$  &  $   1.73$  \\
  1 &  $  2453931.03823$  &  $ -23.27$  &  $   1.64$  \\
  1 &  $  2453931.94201$  &  $ -57.75$  &  $   1.84$  \\
  1 &  $  2453932.03731$  &  $ -55.85$  &  $   1.87$  \\
  1 &  $  2453933.00132$  &  $  -9.75$  &  $   1.84$  \\
  1 &  $  2453933.92609$  &  $  53.85$  &  $   2.32$  \\
  1 &  $  2453934.90529$  &  $  38.18$  &  $   3.75$  \\
  1 &  $  2453934.90853$  &  $  35.21$  &  $   2.09$  \\
  1 &  $  2453960.99790$  &  $  62.44$  &  $   1.87$  \\
  1 &  $  2453961.99776$  &  $   5.91$  &  $   1.83$  \\
  1 &  $  2453963.09343$  &  $ -57.04$  &  $   2.09$  \\
  1 &  $  2454287.93861$  &  $   9.85$  &  $   4.36$  \\
  1 &  $  2454287.94541$  &  $   8.74$  &  $   2.75$  \\
  1 &  $  2454287.95252$  &  $  13.18$  &  $   3.04$  \\
  1 &  $  2454287.96032$  &  $  18.46$  &  $   2.68$  \\
  1 &  $  2454287.96650$  &  $   2.63$  &  $   2.75$  \\
  1 &  $  2454287.97330$  &  $   9.02$  &  $   2.81$  \\
  1 &  $  2454287.97916$  &  $  -1.48$  &  $   4.36$  \\
  1 &  $  2454287.98299$  &  $   8.69$  &  $   3.50$  \\
  1 &  $  2454287.98721$  &  $  13.22$  &  $   3.54$  \\
  1 &  $  2454287.99136$  &  $  20.22$  &  $   3.31$  \\
  1 &  $  2454287.99485$  &  $  15.47$  &  $   3.78$  \\
  1 &  $  2454287.99928$  &  $  15.27$  &  $   4.72$  \\
  1 &  $  2454288.00334$  &  $  20.35$  &  $   4.62$  \\
  1 &  $  2454288.00728$  &  $  22.05$  &  $   3.93$  \\
  1 &  $  2454288.01142$  &  $  17.23$  &  $   3.70$  \\
  1 &  $  2454288.01552$  &  $  15.48$  &  $   3.70$  \\
  1 &  $  2454288.01886$  &  $  11.54$  &  $   3.33$  \\
  1 &  $  2454288.02300$  &  $  23.50$  &  $   3.29$  \\
  1 &  $  2454288.06101$  &  $ -13.13$  &  $   2.45$  \\
  1 &  $  2454288.06483$  &  $ -33.06$  &  $   2.73$  \\
  1 &  $  2454288.06840$  &  $ -27.90$  &  $   2.75$  \\
  1 &  $  2454288.07171$  &  $ -19.28$  &  $   2.48$  \\
  1 &  $  2454288.07475$  &  $ -11.57$  &  $   2.47$  \\
  1 &  $  2454288.07773$  &  $ -11.96$  &  $   2.52$  \\
  1 &  $  2454288.08068$  &  $ -14.12$  &  $   2.42$  \\
  1 &  $  2454288.08364$  &  $ -21.86$  &  $   2.36$  \\
  1 &  $  2454288.08669$  &  $  -6.65$  &  $   2.47$  \\
  1 &  $  2454288.08978$  &  $  -1.64$  &  $   2.57$  \\
  1 &  $  2454288.09297$  &  $  -0.91$  &  $   2.38$  \\
  1 &  $  2454288.09591$  &  $  -4.42$  &  $   2.29$  \\
  1 &  $  2454288.09896$  &  $   4.43$  &  $   2.54$  \\
  1 &  $  2454288.10208$  &  $  -2.81$  &  $   2.57$  \\
  1 &  $  2454288.10554$  &  $   1.99$  &  $   2.29$  \\
  1 &  $  2454288.10893$  &  $  -0.73$  &  $   2.53$  \\
  1 &  $  2454288.11884$  &  $  -5.13$  &  $   2.46$  \\
  1 &  $  2454288.12233$  &  $  -1.83$  &  $   2.54$  \\
  1 &  $  2454288.13167$  &  $ -10.34$  &  $   2.76$  \\
  1 &  $  2454288.13551$  &  $ -10.33$  &  $   3.42$  \\
  1 &  $  2454345.78243$  &  $  20.49$  &  $   2.11$  \\
  1 &  $  2454345.78646$  &  $  20.04$  &  $   2.59$  \\
  1 &  $  2454345.78999$  &  $  12.91$  &  $   2.58$  \\
  1 &  $  2454345.83003$  &  $  13.35$  &  $   2.43$  \\
  1 &  $  2454345.96340$  &  $  13.66$  &  $   2.66$  \\
  1 &  $  2454346.02412$  &  $   8.30$  &  $   2.60$  \\
  1 &  $  2454346.02708$  &  $   9.43$  &  $   2.52$  \\
  1 &  $  2454346.02998$  &  $   6.72$  &  $   2.49$  \\
  1 &  $  2454346.03951$  &  $  14.27$  &  $   2.59$  \\
  1 &  $  2454346.04242$  &  $  18.06$  &  $   2.59$  \\
  1 &  $  2454346.04529$  &  $  17.30$  &  $   2.89$  \\
  1 &  $  2454346.04806$  &  $  18.58$  &  $   2.56$  \\
  1 &  $  2454346.05089$  &  $  16.43$  &  $   2.69$  \\
  1 &  $  2454346.05366$  &  $  12.71$  &  $   2.72$  \\
  1 &  $  2454346.05642$  &  $  37.09$  &  $   2.57$  \\
  1 &  $  2454346.05914$  &  $  16.46$  &  $   2.59$  \\
  1 &  $  2454346.06189$  &  $  21.25$  &  $   2.52$  \\
  1 &  $  2454346.07123$  &  $   5.91$  &  $   2.59$  \\
  1 &  $  2454346.07399$  &  $   7.51$  &  $   2.61$  \\
  1 &  $  2454346.07681$  &  $  15.06$  &  $   2.53$  \\
  1 &  $  2454346.07969$  &  $   0.49$  &  $   2.43$  \\
  1 &  $  2454346.08252$  &  $   0.13$  &  $   2.50$  \\
  1 &  $  2454346.08557$  &  $   7.90$  &  $   2.51$  \\
  1 &  $  2454346.08879$  &  $ -14.99$  &  $   2.54$  \\
  1 &  $  2454346.09189$  &  $  -9.72$  &  $   2.68$  \\
  1 &  $  2454346.09493$  &  $  -3.40$  &  $   2.79$  \\
  1 &  $  2454346.09819$  &  $  -9.89$  &  $   2.73$  \\
  1 &  $  2454346.10137$  &  $ -17.76$  &  $   2.61$  \\
  1 &  $  2454346.10442$  &  $ -18.61$  &  $   2.72$  \\
  1 &  $  2454346.10748$  &  $ -12.82$  &  $   2.76$  \\
  1 &  $  2454346.11065$  &  $ -20.48$  &  $   2.55$  \\
  1 &  $  2454346.11385$  &  $ -24.27$  &  $   2.52$  \\
  1 &  $  2454346.11714$  &  $ -24.44$  &  $   2.62$  \\
  1 &  $  2454346.12030$  &  $ -25.08$  &  $   2.61$  \\
  1 &  $  2454346.12342$  &  $ -21.52$  &  $   2.45$  \\
  1 &  $  2454346.12652$  &  $ -22.09$  &  $   2.61$  \\
  1 &  $  2454346.12959$  &  $ -23.63$  &  $   2.63$  \\
  1 &  $  2454346.14545$  &  $  -2.77$  &  $   2.79$  \\
  2 &  $  2454318.07432$  &  $   2.81$  &  $   7.04$  \\
  2 &  $  2454318.08202$  &  $  17.62$  &  $   7.33$  \\
  2 &  $  2454318.11099$  &  $  24.15$  &  $   6.57$  \\
  2 &  $  2454349.00272$  &  $   6.64$  &  $   5.71$  \\
  2 &  $  2454349.01034$  &  $  13.10$  &  $   6.62$  \\
  2 &  $  2454349.01795$  &  $   2.20$  &  $   4.55$  \\
  2 &  $  2454349.02557$  &  $   5.69$  &  $   5.95$  \\
  2 &  $  2454349.03319$  &  $   9.39$  &  $   4.64$  \\
  2 &  $  2454363.77654$  &  $ -40.68$  &  $   7.64$  \\
  2 &  $  2454363.78414$  &  $ -37.33$  &  $   6.61$  \\
  2 &  $  2454363.90169$  &  $ -32.80$  &  $   6.31$  \\
  2 &  $  2454363.90931$  &  $ -25.78$  &  $   7.03$  \\
  2 &  $  2454363.91692$  &  $ -25.12$  &  $   6.14$  \\
  2 &  $  2454363.92455$  &  $ -33.95$  &  $   8.05$  \\
  2 &  $  2454363.93216$  &  $ -37.52$  &  $   7.64$  \\
  2 &  $  2454363.93978$  &  $ -49.54$  &  $   7.06$  \\
  2 &  $  2454363.94738$  &  $ -48.82$  &  $   6.71$  \\
  2 &  $  2454363.95500$  &  $ -58.17$  &  $   7.52$  \\
  2 &  $  2454363.96261$  &  $ -49.23$  &  $   6.98$  \\
  2 &  $  2454363.97023$  &  $ -66.33$  &  $   7.15$  \\
  2 &  $  2454363.97785$  &  $ -65.61$  &  $   5.92$  \\
  2 &  $  2454363.98547$  &  $ -64.40$  &  $   7.38$  \\
  2 &  $  2454363.99307$  &  $ -61.59$  &  $   7.59$  \\
  2 &  $  2454364.00069$  &  $ -62.49$  &  $   6.41$  \\
  2 &  $  2454364.00830$  &  $ -50.84$  &  $   7.51$  \\
  2 &  $  2454364.01591$  &  $ -44.29$  &  $   7.59$  \\
  2 &  $  2454364.03113$  &  $ -47.85$  &  $   9.03$  \\
  2 &  $  2454364.03874$  &  $ -53.34$  &  $   8.92$  \\
  2 &  $  2454364.04636$  &  $ -56.22$  &  $   7.92$  \\
  2 &  $  2454364.05397$  &  $ -62.86$  &  $   7.99$  \\
  2 &  $  2454364.06158$  &  $ -55.24$  &  $   8.02$  \\
  2 &  $  2454364.07680$  &  $ -56.72$  &  $   8.92$  \\
  2 &  $  2454364.08441$  &  $ -52.38$  &  $   8.53$  \\
  3 &  $  2453897.11172$  &  $  -2.89$  &  $   5.00$  \\
  3 &  $  2453899.12580$  &  $  52.53$  &  $   4.77$  \\
  3 &  $  2453900.11917$  &  $ -34.51$  &  $   4.65$  \\
  3 &  $  2453901.10176$  &  $ -37.51$  &  $   4.88$
\enddata

\tablenotetext{a}{ (1) HIRES, Keck~I 10m telescope, Mauna Kea, Hawaii.
  (2) HDS, Subaru 8m telescope, Mauna Kea, Hawaii, based on
  observations and data reduction procedures described in this work.
  (3) HDS, Subaru 8m telescope, Mauna Kea, Hawaii, based on
  observations and data reduction procedures described by Bakos et
  al.~(2007). }

\end{deluxetable}
\begin{deluxetable}{llr}

\tablecaption{Photometric Measurements of HAT-P-1\label{tab:phot}\tablenotemark{a}}
\tablewidth{0pt}
\tablehead{
\colhead{Telescope Code\tablenotemark{b}} &
\colhead{Heliocentric Julian Date} &
\colhead{Relative Intensity} \\
}
\startdata
  $  1$  &  $  2454381.71537$  &  $  0.9988$ \\
  $  1$  &  $  2454381.71588$  &  $  1.0002$ \\
  $  1$  &  $  2454381.71640$  &  $  1.0007$ \\
  $  1$  &  $  2454381.71691$  &  $  1.0028$ \\
  $  1$  &  $  2454381.71904$  &  $  1.0018$ \\
  $  1$  &  $  2454381.71955$  &  $  1.0002$
\enddata

\tablenotetext{a}{Full table available in online addition.}
\tablenotetext{b}{
(1) Nickel telescope, Mt. Hamilton, California.
(2) MAGNUM 2m telescope, Haleakala, Hawaii.}

\end{deluxetable}
\begin{deluxetable}{cc}

\tablecaption{Midtransit times of HAT-P-1\label{tab:tc}}
\tablewidth{0pt}

\tablehead{
\colhead{Midtransit time~[HJD]} &
\colhead{Reference}
}

\startdata
$2453984.39700 \pm 0.00900$  &  1 \\
$2453979.92994 \pm 0.00069$  &  2\tablenotemark{a} \\
$2453988.86197 \pm 0.00076$  &  2 \\
$2453997.79200 \pm 0.00054$  &  2 \\
$2453997.79348 \pm 0.00047$  &  2 \\
$2454006.72326 \pm 0.00059$  &  2 \\
$2454015.65338 \pm 0.00107$  &  2\tablenotemark{a} \\
$2454069.23795 \pm 0.00290$  &  2 \\
$2454363.94601 \pm 0.00091$  &  3 \\
$2454381.80849 \pm 0.00125$  &  3 
\enddata
\tablecomments{References:~~(1) Bakos et al.~(2007), (2) Winn et
al.~(2007), (3) This work.}

\tablenotetext{a}{The midtransit time for this event that was
  originally reported by Winn et al.~(2007) was incorrect. The data
  reported here have been corrected.}

\end{deluxetable}

\begin{deluxetable}{lcc}

\tablecaption{System Parameters of HAT-P-1\label{tab:params}}
\tablewidth{0pt}

\tablehead{
\colhead{Parameter} & \colhead{Value} & \colhead{Uncertainty}
}

\startdata
{\it Transit ephemeris:} & &  \\
Orbital period, $P$~[d]                             & $4.4652934$     & $0.0000093$ \\
Midtransit time~[HJD]                               & $2454363.94656$ & $0.00072$   \\
& &  \\
{\it Rossiter-McLaughlin parameters:} & &  \\
Projected spin-orbit angle, $\lambda$~[deg]         & $3.7$           & $2.1$   \\
Projected stellar rotation rate, $v\sin i_*$~[km~s$^{-1}$] & $3.75$    & $0.58$  \\
& &  \\
{\it Photometric transit parameters:} & &  \\
Planet-to-star radius ratio, $R_p/R_\star$           & $0.11295$      &  $0.00073$  \\
Orbital inclination, $i$~[deg]                     &  $86.28$        &  $0.20$     \\
Scaled semimajor axis, $a/R_\star$                   &  $10.67$       &  $0.25$     \\
Transit impact parameter                            &   $0.693$       &  $0.023$    \\
Transit duration~[hr]                               &   $2.798$       &  $0.019$    \\
Transit ingress or egress duration~[hr]             &   $0.510$       &  $0.031$    \\
& &  \\
{\it Spectroscopic orbital parameters:} & &  \\
Orbital eccentricity, $e$                           & $0$       &  assumed\tablenotemark{b}  \\
Velocity semiamplitude, $K$~[m~s$^{-1}$]             & $59.3$    &  $1.4$ \\
Additive velocity (Keck/HIRES)~[m~s$^{-1}$]          & $1.26$     &  $0.66$ \\
Additive velocity (Subaru, this work)~[m~s$^{-1}$]   & $-45.3$   &  $1.5$ \\
Additive velocity (Subaru, Bakos et al.~2007)~[m~s$^{-1}$] & $13.0$ & $3.1$ \\
& &  \\
{\it Derived system parameters:} & &  \\
$M_\star$~[M$_\odot$]\tablenotemark{b}              &  $1.133$   &  $0.077$  \\
$R_\star$~[R$_\odot$]\tablenotemark{c}              &  $1.115$   &  $0.050$ \\
$M_p$~[M$_{\rm Jup}$]\tablenotemark{c}              &  $0.524$   &  $0.031$  \\
$R_p$~[R$_{\rm Jup}$]\tablenotemark{c}              &  $1.225$   &  $0.059$ \\
$M_p/M_\star$ \tablenotemark{c}                    &  $0.000441$ &  $0.000020$
\enddata

\tablenotetext{a}{When the orbital eccentricity is taken to be a free
  parameter, we find an upper limit of 0.067 with 99\% confidence.}

\tablenotetext{b}{The stellar mass was not determined from our
  data. The value given here is from Torres, Winn, \& Holman (2008).}

\tablenotetext{c}{This result depends on the assumed value and
  uncertainty in the stellar mass.}

\end{deluxetable}

\clearpage

\end{document}